\documentclass{article}
\usepackage{spconf,amsmath,amssymb,graphicx,hyperref}
\usepackage{booktabs, multirow}
\title{Advancing speech summarization in Multi-modal LLMs with Reinforcement Learning}
\name{Shaoshi Ling, Gang Liu, Guoli Ye, Jinyu Li}
\address{Microsoft CoreAI, USA}
\begin{document}
\ninept
\maketitle
\begin{abstract}
%Speech summarization (SSum) is a critical component of spoken content understanding, particularly in the era of rapidly growing spoken and audiovisual data. Recent advances in multi-modality large language models (MLLMs) have enabled leveraging the power of LLMs to generate textual summaries directly from speech with out relying on intermediate transcriptions. These models have even demonstrated zero-shot capabilities, performing end-to-end summarization across arbitrary domains, producing controllable styles, and remaining accessible with minimal setup. Despite these advantages open-source MLLMs s till exhibit a noticeable performance gap compared to LLMs in text modality, posing challenges for production scale deployment. In this work, we present several approaches to enhance the summarization capabilities in MLLMs. Our resulting model achieves substantial improvements and closes the gap with state-of-the-art text-based LLMs.
Speech summarization (SSum) is a critical component of spoken content understanding, particularly in the era of rapidly growing spoken and audiovisual data. Recent advances in multi-modal large language models (MLLMs), leveraging the power of LLMs, enable generating textual summaries directly from speech without intermediate transcriptions, while supporting controllable styles and zero-shot generalization. However, open-source MLLMs continue to lag behind the state-of-the-art text-based LLMs, limiting their practical deployment for speech summarization. In this work, we present a novel multi-stage reinforcement learning (RL) training framework to enhance the speech summarization capabilities in MLLMs. Our model delivers substantial improvements over strong baselines, outperforms much larger MLLMs, and significantly narrows the gap with state-of-the-art text-based LLMs.
\end{abstract}
\section{Introduction}
\label{sec:intro}

As audio and audiovisual content increasingly dominate modern communication and media consumption, speech summarization (SSum)—the task of generating concise and coherent textual summaries directly from spoken input—has become a critical capability for improving accessibility, productivity, and information retrieval. By enabling faster access to information, supporting academic and business workflows, and facilitating personal consumption, SSum plays a pivotal role in bridging the gap between the vast amount of spoken data and the convenience of text-based interaction. Its importance continues to grow as society produces large amounts of spoken and audiovisual data across meetings, lectures, podcasts, and social media platforms.

Traditional SSum has been approached through a cascaded pipeline, where automatic speech recognition (ASR) is followed by text summarization\cite{retkowski2025speech}. While effective to some extent, this approach introduces error propagation from ASR and often struggles to preserve nuances such as discourse structure, prosody, and emphasis. To address these shortcomings, end-to-end speech summarization methods have been proposed \cite{matsuura2023leveraging, matsuura2023transfer, sharma2022end, shang2024end, kang2024prompting, eom2024squba}, which generate summaries directly from speech without an intermediate transcription step. These approaches typically combine a speech encoder with an independently trained LLM or a task-specific summarization module, thereby aiming to reduce the error propagation introduced by ASR. However, because the components are independently trained and their finetuning is often constrained by limited training data, these systems generally lack strong instruction following ability, zero-shot generalization, and controllability\cite{retkowski2025speech}.

%For the first time, MLLMs make end-to-end SSum accessible with minimal setup.
Building on the success of Large Language Models (LLMs), Multi-modal Large Language Models (MLLMs) extend LLMs to handle multi-modal inputs, integrating modalities beyond text. This extension is particularly critical for speech summarization, where the input is inherently multi-modal: acoustic signals encode not only linguistic content but also paralinguistic information such as speaker emphasis, prosody, and emotion. Leveraging such information has the potential to produce summaries that are more accurate and contextually faithful compared with text-only models. Commercial models such as GPT-4o-Audio \cite{hurst2024gpt} and Gemini-2.5 \cite{comanici2025gemini} have demonstrated promising SSum capabilities, but their large size, closed-source nature, and limited accessibility make them challenging to deploy at scale for general users. In contrast, open-source models such as Qwen2-Audio \cite{chu2024qwen2} have been applied to zero-shot SSum without task-specific training. Similarly, Phi-4MM \cite{abouelenin2025phi} is the first open-source model with general-purpose speech summarization capability that supports long-form audio inputs up to 2 hours. However, these open-source models still exhibit a substantial performance gap compared with the state-of-the-art commercial models like GPT-4o \cite{hurst2024gpt}. Moreover, all current MLLMs consistently underperform in the audio modality relative to the text modality, highlighting a persistent modality gap that limits their effectiveness for speech summarization.

%To this end, we propose a three-stage training framework to enhance the SSum capability of MLLMs. 
To this end, we propose the following approach. We first construct large-scale synthetic datasets to improve the model’s ability to follow a wide range of instructions, thus strengthening its general instruction following capability. Next, we employ large-scale reinforcement learning (RL) of on-policy knowledge distillation \cite{gu2023minillm, agarwal2024policy, cao2025llm} to transfer knowledge from powerful text-based LLMs into student MLLM, thereby reducing the modality gap. Finally, we bootstrap the model with another RL method Direct Preference Optimization (DPO) \cite{rafailov2023direct}, which mitigates issues such as hallucinations and improves robustness. 
%Together, these stages yield models that deliver significant improvements over strong baselines, outperform much larger commercial MLLMs, and substantially narrow the gap with state-of-the-art text-based LLMs. 

Our contributions can be summarized as follows:
\begin{itemize}
\item \textbf{Novel training framework}: We propose a novel multi-stage RL training framework that improves instruction-following, modality alignment, and cross-lingual generalization for speech summarization in MLLMs.
\item \textbf{On-policy knowledge distillation}: We propose an on-policy knowledge distillation method that enables effective knowledge transfer from large text-based LLMs into the audio-conditioned MLLMs.
\item \textbf{Strong empirical performance}: Our model achieves up to a 28\% relative improvement over competitive baselines, surpasses much larger MLLMs such as GPT-4o-audio, and further narrows the performance gap with state-of-the-art LLMs.
\end{itemize}

%Under this framework, the resulting model narrows the gap between text and audio modalities, achieving strong performance in summarizing both spoken content and textual documents across diverse domains and languages. Moreover, it can generate summaries in different styles by varying the instruction.

%Our experiments demonstrate that the proposed framework substantially improves speech summarization (SSum) quality in multi-modal large language models (MLLMs). Compared to strong baselines, our approach achieves up to a 21\% relative improvement, highlighting its effectiveness across arbitrary domains with strong instruction-following capabilities. We further observe zero-shot cross-lingual generalization, where the model improves summarization quality in languages unseen during training. 
%Our paper makes main contributions as follows:
%\begin{itemize}
%\item We propose a novel multi-stage training framework combining synthetic instruction tuning, on-policy knowledge distillation, and DPO.
%\item We introduce an on-policy knowledge distillation method for cross-modal distillation and effectively narrows the gap between speech and text summarization.
%\item Our empirical results achieves up to a 21\% relative improvement over baselines and narrows the performance gap with state-of-the-art models that are substantially larger in size.
%\end{itemize}

\begin{figure*}[pht]
\centering
{\includegraphics[width=1.0\textwidth,height=0.18\textwidth]{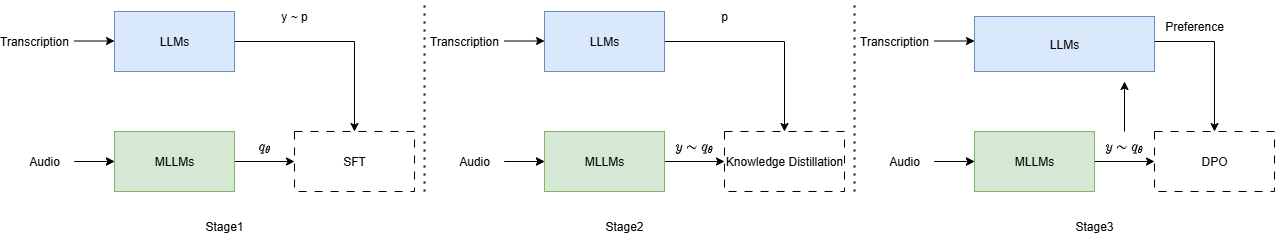}} %
\caption{The overview of three-stage training process.}
\label{fig:overview}
\end{figure*}

\section{Method}
We propose a three-stage training framework designed to enhance the speech summarization capabilities of multi-modal large language models (MLLMs). As illustrated in Figure \ref{fig:overview}, the framework consists of: (1) Supervised finetuning (SFT) on synthetic data. (2) On-policy knowledge distillation (KD) to transfer summarization ability from strong text-based LLMs. (3) Direct Preference Optimization (DPO) to mitigate hallucinations.
\subsection{Synthetic data for SFT}
\label{sec:sft}
Supervised fine-tuning (SFT) serves as the foundation of our multi-stage training framework because it directly shapes the model’s ability to follow instructions and generate useful summaries. To further enhance the instruction-following capabilities of MLLMs, we construct a large and diverse synthetic dataset oriented toward summarization.

Our data collection process builds on the approach introduced in Phi-4MM \cite{abouelenin2025phi}, where anonymized audio recordings are paired with their transcripts. The audio spans a broad range of topics, ensuring representation of both everyday and domain-specific speech. While the original Phi-4MM pipeline generated ten query–summary pairs per audio sample in a single pass, we extend and refine this process to further improve diversity and quality. Specifically, for each transcript, we first use GPT-4.1 to generate multiple candidate queries, each accompanied by an importance score reflecting its suitability for summarization. Queries with low scores are discarded to retain only high-quality instructions. From the filtered set, we then randomly select one query to associate with the corresponding audio, emphasizing diversity while avoiding redundancy. The resulting queries vary in granularity and style, ranging from concise requests for short summaries to more elaborate instructions requiring structured outputs such as bullet points, JSON objects, or email-style narratives. Given the selected query and transcript, GPT-4.1 is then prompted to generate a reference summary. This two-step design—importance scoring followed by query selection—ensures that the resulting dataset balances diversity with consistency and relevance. Compared with the baseline pipeline in Phi-4MM \cite{abouelenin2025phi}, where GPT-4 was instructed to generate ten query–summary pairs for each audio in a single run, our method produces summaries that are significantly longer, richer in content, and more varied in format.

%Specifically, we first use GPT-4.1 to generate multiple queries along with an importance score indicating the suitability of each query for summarization. Data with low scores are discarded. We randomly sampled one query from queries generated.These queries are designed to capture both specific and general aspects of the conversation and vary in format, including length (number of words or sentences) and structure (e.g., bullet points, JSON, or email). Given the selected query and the corresponding transcript, we then use GPT-4.1 to construct the summary. 
%We This pipeline produces higher-quality and more diverse query–summary pairs compared to the method used in Phi-4MM~\cite{abouelenin2025phi}.

\subsection{On-policy knowledge distillation}
\label{sec:KD}
Knowledge distillation (KD) \cite{li2014learning, hinton2015distilling} is a widely used technique for transferring the behavior of a powerful teacher model to a smaller student model. However, in the case of speech summarization, direct distillation is limited by a strong distributional mismatch: the large teacher model produces outputs in the text modality, often with more expressive linguistic patterns than the audio-conditioned student can reproduce, and the student is trained only to imitate teacher-generated sequences rather than to correct its own errors. This mismatch leads to instability and weaker generalization, especially when the student encounters inputs that deviate from the teacher’s training distribution. This phenomenon is commonly referred to as mode collapse in the literature \cite{cao2025llm}.

To overcome these issues, we adopt an on-policy distillation strategy, where the student learns from its own generated sequences rather than from teacher outputs alone. During training, the student generates rollouts conditioned on audio input, and the teacher provides token-level probability supervision on these sequences. In this way, the supervision remains grounded in the student’s trajectory while still benefiting from the teacher’s richer linguistic knowledge. This reduces exposure bias and enables the student to improve on the kinds of errors it is likely to make at inference time, leading to more stable optimization and closer alignment between capacity and supervision. Importantly, this design also facilitates cross-modality transfer: although the teacher operates in the text modality, the student effectively absorbs the teacher’s summarization ability while grounding it in audio, thereby narrowing the performance gap between modalities. Moreover, when using on-policy data during distillation, the student receives token-specific feedback from the teacher’s logits on the erroneous tokens in its self-generated output sequences. This enables a form of feedback loop akin to what is observed in RL, which helps minimize the train-inference distribution mismatch. 

%the complexity of the output space poses unique challenges. Large teacher LLMs, operating in the text modality, often capture richer modes of expression than a student MLLM can reproduce due to its limited capacity and the additional difficulty of handling audio inputs. Directly training the student on teacher outputs therefore risks mismatch and instability.

%To address this issue, we adopt an on-policy distillation strategy, where the student learns from sequences it generates during training rather than solely from teacher-produced text. This approach encourages stability and improves alignment between the student’s capacity and the supervision it receives. Building on previous work\cite{gu2023minillm, agarwal2024policy}, we extend this method to enable cross-modality transfer\cite{ling2025customizing}, distilling knowledge from a text-based teacher LLM into an audio-conditioned MLLM.

%To alleviate this problem, many approach has been proposed to train on its self-generated sequences that are on-policy \cite{gu2023minillm, agarwal2024policy}. We adopted those approaches and extend this method to distilliation from text modaility to audio modality \cite{ling2025customizing} so that the MLLMs can learn from larger LLMs. Specifically, the on-policy loss L is given by:

Formally, given an audio input $x$, the student MLLMs $P_S^\theta$ produces an output sequence $y \sim p_S(\cdot|x)$. For each token in the generated sequence, the student imitates the token-level probability distribution of the teacher LLMs $P_T$ . The training objective is expressed as:
\begin{equation}
L(\theta) = \mathbb{E}_{x \sim X} \Bigg[ 
    \mathbb{E}_{y \sim p_S(\cdot|x)} \Big[ 
        \mathcal{D}_{KL}\!\left( p_T \,\|\, p_S^\theta \right)(y|x) 
    \Big] 
\Bigg]
\end{equation}

%Given an audio input x, the student MLLMs $P_S^\theta$ generates the output sequence y and imitates the teacher LLMs $P_T$ token-level distributions, $p_S^\theta(y|x)$. We adopt the same strategy in \cite{agarwal2024policy} that we do not backpropagate through the student’s sampling distribution $p_S^\theta(\cdot|x)$, similar to on-policy imitation. Not backpropagating through the sampling makes the training stable and computationally efficient. The gradient can be written as:
Importantly, we follow the strategy of \cite{agarwal2024policy} and do not back-propagate through the sampling distribution $p_S^\theta(\cdot|x)$. This design choice avoids high-variance gradient estimates, yielding stable and computationally efficient training. The gradient of the objective can then be written as:
 \begin{align}
\nabla_\theta L(\theta) 
&= \mathbb{E}_{x \sim X} \, \mathbb{E}_{y \sim p_S(\cdot|x)} 
\Bigg[ \sum_{z} p_T(z|x,y) \, \nabla_\theta \log p_S^\theta(z|x,y) \Bigg]
\end{align}
where $z$ are tokens in the vocabulary. Note that this equation has the same form as the policy gradient approach \cite{sutton1999policy}. In our formulation, the reward is given by $p_T(z|x,y)$ which provides the feedback for how good the token $z$ is given context $x,y$.

Compared with traditional knowledge distillation where the teacher generates the sequence, the supervision is more consistent with the student’s actual behavior, leading to improved stability, more natural generalization and reduced risk of mode collapse. In practice, this stage produces the largest performance gains in our framework, demonstrating that on-policy distillation is particularly effective for transferring the linguistic competence of a text-only LLM into an audio-conditioned MLLM, while simultaneously narrowing the modality gap and enhancing robustness.

%By leveraging supervision from a stronger text-based LLM teacher, this stage substantially reduces the performance gap between audio and text modalities while maintaining robustness against instability and mode collapse.

\subsection{DPO}
\label{sec:dpo}
While supervised fine-tuning and on-policy knowledge distillation substantially enhance the SSum of the model, we observe that they also introduce undesirable artifacts. In particular, the student MLLM sometimes produces degenerate outputs that exploit weaknesses in the training objective—for example, generating repetitive phrases or hallucinations—that nevertheless receive high scores from the teacher model. This phenomenon, commonly referred to as reward hacking in the RL literature \cite{weng2024rewardhack}, undermines the reliability of the system.

To address this issue, we introduce a final stage of training based on Direct Preference Optimization (DPO) \cite{rafailov2023direct}. DPO leverages pairwise preference data to align the model more closely with human-like quality judgments. For each input audio $x$, we sample two distinct hypotheses from the student model. These hypotheses are then evaluated by GPT-4.1, which expresses a preference between them. The preferred response, denoted $y^+$, typically represents a coherent and well-structured summary, while the less preferred response, denoted $y^-$, often contains degenerate behaviors such as repetition or hallucination. Instances where GPT-4.1 expresses no clear preference are discarded. %Repeating this process offline yields a preference dataset $D = \{X, a^+, a^-\}_n$.
%We observe reward hacking when training in the previous stages. Student MLLMs sometimes produces degenerated sentencesy that receive high scores from the teacher LLMs (e.g., repeated phrases, wrong formats) during sampling. Thus we apply the third stage DPO \cite{rafailov2023direct} to fix this issue. The DPO algorithm relies on pair-wise preference data. Specifically, two distinct hypotheses are independently sampled from the policy MLLMs model. These responses are then ranked based on GPT-4.1 preferences, resulting in a preferred response $a^+$ containing regular hypothesis and a less preferred one $a^-$ containing the hypothesis with the reward hacks. If model does not have clear preference, the pairs of data will be discarded. This process is repeated multiple times offline to construct the preference dataset $D = \{X, a^+, a^-\}_n$.

DPO then optimizes the following objective function:
\begin{equation}\label{attention}
  \begin{aligned}
L_{DPO}=\mathbb{E}_{x \sim X}\Bigg[log\sigma(\beta log \frac{\pi(y^+|x)}{\pi_{ref}(y^+|x)} - \beta log \frac{\pi(y^-|x)}{\pi_{ref}(y^-|x)})\Bigg]
\end{aligned}
\end{equation}
where $\pi$ denotes the current model which is initialized from  the reference model, $\pi_{ref}$, which is the checkpoint obtained from on-policy knowledge distillation in section \ref{sec:KD}. By applying DPO as the final step, we effectively reduce hallucinations and improve the overall consistency of generated summaries. 
%This formulation encourages the model to increase the relative likelihood of preferred outputs while suppressing degenerate alternatives, thereby correcting biases introduced in earlier stages.

\section{Experimental Settings}
\subsection{Baseline}
\label{sec:baseline}
The baseline model builds on Phi-4MM \cite{abouelenin2025phi}, which was trained on 500k summarization samples consisting of 50k audio recordings, each paired with 10 query–summary pairs. For clearer analysis and comparison, we do not use the final publicly released model. Instead, we build our model on the pre-training stage model, where it was trained on large-scale automatic speech recognition (ASR) data to align the audio encoder with text-based Phi-4Mini \cite{abouelenin2025phi} in the semantic space.  

\subsection{Training Datasets}
In the SFT stage, we expand the audio data from 50k to 1M samples compared to the baseline. For each audio, we generate only a single query–response pair, as described in \ref{sec:sft}. Compared with the baseline, our synthetic pipeline produces queries that are more diverse and summaries that are substantially richer in content. On average, summaries are three times longer than those in the baseline dataset, capturing finer details. 

For the knowledge distillation and DPO stages, we construct data using the same process but omit the summary generation step. We sample 35k high-quality audios with query for training. All of the above data are in English only.
%With improvements in the synthetic model, data generation pipeline, and filtering, the quality of our SFT data is substantially higher than that of the baseline. The queries are more diverse, and the summaries are more informative, with the average word count being three times higher than in the baseline. In the knowledge distillation and DPO stages, we follow the same data construction process as in SFT, except we omit the summary generation step. We sample 35k high-quality audios with query for training.

\subsection{Training Setting}
% using the AdamW optimizer with a peak learning rate of 1e-4. A linear warmup–decay schedule is applied (1,000 warmup steps, 40,000 total steps)
In all training stages, the audio encoder is kept frozen; only the audio projector and LoRA modules are updated. The LoRA configuration uses $\alpha=32$ and $rank=16$. During the SFT stage, the model is trained for 2 epochs on 32 A100 GPUs. For the knowledge distillation stage, we use the GPT-4o text only mode \cite{hurst2024gpt} as the teacher model. Training is conducted with the verl framework \cite{sheng2025hybridflow} where vLLM \cite{kwon2023efficient} is used for rollout. The student model is trained on 8 A100 GPUs not including the GPUs allocated to the teacher model. In the DPO stage, the model is trained for 1 epoch on 32 A100 GPUs. Across all stages, training is performed on audio inputs up to 30 minutes in length (corresponding to approximately 22.5k tokens). Given the 128k context length of the language decoder, our model supports up to 2.8 hours of audio during inference.

\subsection{Evaluation Setting}
We evaluate speech summarization performance on three benchmarks: %an in-house dataset (Golden3), a public benchmark (AMI \cite{carletta2005ami}) and a multilingual benchmark (FLORAS \cite{chen2024floras}).

\begin{itemize}
\item Golden3: An in-house meeting dataset with 108 recordings (average 6 minutes each) and 321 instructions. The data is English-only and covers diverse topics.
\item AMI\cite{carletta2005ami}): A public English-only meeting corpus (~100 hours) with multimodal streams. We evaluate on the test split (20 meetings, ~32 minutes each) using close-talking audio. Each meeting includes 3 summarization instructions (60 total).
\item FLORAS\cite{chen2024floras}: is a multilingual benchmark designed to assess model performance on raw, long-form conversational audio from YouTube. We selected 548 recordings ranging from 5 minutes to 1 hour in duration, spanning languages including Spanish, Italian, French, German, Portuguese, Chinese, and Japanese.
\end{itemize}
%and performs speech summarization directly
During inference, our model processes long-form audio in one shot without segmentation. For evaluation metrics, we follow the same setting as in Phi-4MM\cite{abouelenin2025phi} with small modifications. The outputs are scored by GPT-4.1 against the transcription for overall quality. The overall quality score, ranging from 1 to 7, measures accuracy in capturing details, coherence, writing style, degree of hallucination, and adherence to instruction-specific requirements regarding format, content, and length. 

\begin{table}[h!]
\centering
\begin{tabular}{lccc}
\toprule
 %& \textbf{Golden3}  & \textbf{AMI} & \textbf{Floras} \\
  &\textbf{Golden3$\uparrow$} & \textbf{AMI$\uparrow$} & \textbf{Floras$\uparrow$} \\
\midrule
GPT-4o\cite{hurst2024gpt} Audio& 6.26 & 5.83 & 5.77\\
GPT-4o\cite{hurst2024gpt} Text& 6.57 & 6.75 & 6.82 \\
Phi-4MM\cite{abouelenin2025phi} Text& 5.50 & 5.28 &5.17\\
Phi-4MM\cite{abouelenin2025phi} Audio& 5.02 & 4.55 & 4.69\\
%Gemini2.5-flash\cite{comanici2025gemini} Audio & 6.54& 6.70 & 6.75\\
\midrule
Phi-4MM replicated	& 4.84	& 4.13	& 4.16 \\
Phi-4MM SFT	& 4.97	& 5.14	& 5.14 \\
Phi-4MM SFT+KD&6.05 &5.75 & 4.93\\	
Phi-4MM SFT+KD+DPO & 6.36 & 6.26& 5.74			\\

\bottomrule
\end{tabular}
\caption{Summarization results}
\label{tab:main_results}
\end{table}

\section{Experimental Results}
\subsection{Main Results}
Table~\ref{tab:main_results} presents results across the three benchmarks, comparing our proposed PhiSSum models with both open-source baselines and state-of-the-art text-based and multi-modal systems. To establish a fair baseline, we replicated the Phi-4MM pre-training stage using 500k summarization samples restricted to summarization-only data, whereas the released Phi-4MM was trained on additional spoken QA data. As expected, this replication underperforms the released model.
%presents the performance of our models across the three evaluation benchmarks. We compare against several baselines, including text-based summarization models that take the ground-truth transcription as input. The second half of the table reports the results from our proposed models, denoted as Phi-4MM which are trained on top of the Phi-4MM backbone at its pre-training stage (see\ref{sec:baseline}). 

%We begin by replicating the Phi-4MM model using the same 500k summarization dataset as the baseline. It is important to note that the official open-sourced Phi-4MM model was trained on a much larger dataset that also included supplementary tasks such as spoken question answering (SQA), which provide additional supervision for summarization. This explains why the open-sourced Phi-4MM outperforms our replication. To ensure fairness, however, we restrict our replication to summarization-only data.

Our first-stage model, Phi-4MM-SFT, already delivers substantial improvements over the replicated Phi-4MM, demonstrating the effectiveness of the enhanced synthetic data pipeline. Building upon this, we apply on-policy knowledge distillation (KD) with GPT-4o (text mode) as the teacher. This stage yields the most substantial single boost in performance, with notable improvements on Golden3 and AMI. The results confirm that transferring summarization ability from a powerful text-only teacher into an audio-conditioned student is highly effective. However, this stage also introduces more frequent hallucinations as mentioned in section \ref{sec:dpo}. For instance, given the query "List all the TV shows specifically named in this conversation", the model outputs "The TV shows specifically mentioned in the conversation are: - How to Get Away - How to Get Away ...." repeating "How to Get Away" multiple times in succession. 

To mitigate these issues, we incorporate a third stage of training with Direct Preference Optimization (DPO). DPO effectively reduces hallucinations and improves overall reliability by aligning the model’s outputs with human-preferred responses. As a result, our final model significantly outperforms the baseline, achieving a 28\% relative improvement and surpassing GPT-4o-audio, despite being much smaller in scale, while also narrowing the performance gap with the state-of-the-art LLMs. 

Another notable finding is the model’s strong generalization across languages, even though training used only English data. On the multilingual Floras benchmark, our model maintains performance close to GPT-4o-Audio, demonstrating robust zero-shot cross-lingual transfer. These results highlight that, with carefully staged training—including synthetic data, on-policy knowledge distillation, and preference alignment—smaller open-source models can rival or even surpass much larger commercial systems in speech summarization.

%Furthermore, although trained exclusively on English data, the model generalizes strongly to multilingual evaluation in Floras, demonstrating promising zero-shot cross-lingual capability.
%We then applied on-policy knowledge distillation (KD) on top of Phi-4MM-SFT, which provided the largest performance gains across all three evaluation stages. Here, the teacher model for KD was the GPT-4o text model. However, after this stage, our error analysis revealed notable hallucinations, which we suspect were introduced during KD. To mitigate this issue, we further applied DPO training.

%Our final model significantly outperforms the Phi-4MM baseline, and its performance approaches that of GPT-4o Audio, despite being much smaller in size. Furthermore, our model exhibits zero-shot cross-lingual capabilities: although trained exclusively on English data, it performs strongly on multilingual evaluation sets such as Floras.

\begin{table}[h!]
\centering
\begin{tabular}{lccc}
\toprule
 %& \textbf{Golden3} & \textbf{AMI} & \textbf{Floras} \\
 &\textbf{Golden3$\uparrow$} & \textbf{AMI$\uparrow$} & \textbf{Floras$\uparrow$}\\
\midrule
Phi-4MM replicated	& 4.84	& 4.13	& 4.16 \\
Phi-4MM SFT	100k & 4.82 & 4.83 & 4.63 \\
Phi-4MM SFT	200k & 4.97 & 4.98 & 4.76 \\
Phi-4MM SFT	400k & 5.03 & 5.00 & 4.97 \\
Phi-4MM SFT	600k & 4.97 & 5.05 & 5.04 \\
Phi-4MM SFT	1M   & 4.97 & 5.14 & 5.14 \\
\bottomrule
\end{tabular}
\caption{Summarization results on different size of SFT datasets}
\label{tab:sft_results}
\end{table}

\subsection{Ablation Study}
Table~\ref{tab:sft_results} shows that larger SFT datasets consistently improve performance. Even 200k examples surpass the 500k data in the baseline, underscoring the importance of data quality. The 1M-sample set yields the strongest results. 

Table~\ref{tab:ablation} compares different teacher models for on-policy knowledge distillation. When using Phi-4MM text mode as the teacher, one might expect results similar to Phi-4MM since they share the same backbone. However, the teacher provides modest gains on Golden3 but leads to degradation on AMI and FLORAS, which we attribute to reward hacking behaviors more prevalent in smaller LLMs. In contrast, using GPT-4o (text mode) as the teacher yields consistent improvements across all datasets, validating the importance of leveraging a stronger teacher to mitigate hallucinations and achieve improvements across all datasets. 

We also attempted to apply DPO directly to the model in Table~\ref{tab:ablation}. Both DPO and on-policy KD are recognized as state-of-the-art RL methods, but their effectiveness varies across datasets. On Golden3, KD substantially outperforms DPO, while on AMI and FLORAS, DPO yields stronger results. Our error analysis suggests that KD produces higher-quality summaries overall, but the hallucinations lower evaluation scores. This complementary behavior underscores the necessity of combining KD with DPO in our framework.
%reports results when varying the size of the SFT dataset. As expected, increasing the number of high-quality synthetic examples consistently improves performance. Notably, even a 200k-sample dataset already outperforms the baseline which used 500k data, highlighting the effectiveness of our improved data generation and filtering process. 
%presents the performance comparison across different sizes of SFT data. As expected, larger SFT datasets lead to better model performance. Notably, the 200k high-quality SFT dataset already outperforms the 500k open-sourced SFT dataset.

%In particular, we experimented with using the Phi-4mini text model as the teacher. Since Phi-4MM and Phi-4mini share the same backbone, one might expect similar results; however, as shown in Table~\ref{tab:main_results}, we observed a performance gap depending on the modality of the input. Specifically, we observed performance gains on Golden3 but degradation on AMI and Floras, which we attribute to reward hacking—a common issue, especially for smaller LLMs. In contrast, larger LLMs such as GPT-4o mitigate these hallucination problems and achieve improvements across all datasets.

%Both DPO and on-policy KD are considered state-of-the-art RL algorithms in the literature. Our results show that on Golden3, on-policy KD substantially outperforms DPO, while the opposite holds for AMI and Floras. After further error analysis, we believe this is because on-policy KD produces higher-quality summaries overall, but the increased hallucinations lower the evaluation scores on those two datasets.

\begin{table}[h!]
\centering
\begin{tabular}{lccc}
\toprule
 %& \textbf{Golden3} & \textbf{AMI} & \textbf{Floras} \\
 &\textbf{Golden3$\uparrow$} & \textbf{AMI$\uparrow$} & \textbf{Floras$\uparrow$}\\
\midrule
Phi-4MM & 5.02&4.55&4.69\\
\midrule
Phi-4MM + DPO & 5.53&5.2&5.36\\
\midrule
Phi-4MM + KD(Phi-4MM text) &5.36&4.46&4.13\\
Phi-4MM + KD(GPT-4o) & 6.03 &	5.32&4.84 \\
\bottomrule
\end{tabular}
\caption{Comparison of different teachers in on-policy KD}
\label{tab:ablation}
\end{table}

\section{Conclusion}
In this work, we introduced a multi-stage training framework designed to enhance the speech summarization capabilities of multi-modal large language models (MLLMs). By combining supervised fine-tuning on synthetic data, on-policy knowledge distillation from a stronger text-based teacher, and Direct Preference Optimization, our approach delivers consistent and substantial improvements. The resulting model achieves up to a 28\% relative improvement over strong baselines, outperforms considerably larger state-of-the-art MLLMs such as GPT-4o-audio, and narrows the performance gap with leading text-based LLMs. Future directions include incorporating speaker-aware summarization and leveraging time-aligned information to improve temporal coherence in speech summarization. %We believe these extensions will further advance the robustness, fidelity, and applicability of MLLMs for real-world speech summarization tasks.

%In this work, we introduced a multi-stage training framework designed to enhance the speech summarization capabilities of MLLMs. Our model achieves up to a 28\% relative improvement over strong baselines and outperforms considerably larger state-of-the-art MLLMs such as GPT-4o-audio. Nevertheless, due to its limited size, our model still lags behind GPT-4o and exhibits a higher tendency to generate hallucinations. Future directions include incorporating speaker-aware summarization and leveraging time-aligned information for improved temporal coherence.

\vfill\pagebreak

% References should be produced using the bibtex program from suitable
% BiBTeX files (here: strings, refs, manuals). The IEEEbib.bst bibliography
% style file from IEEE produces unsorted bibliography list.
% -------------------------------------------------------------------------
\bibliographystyle{IEEEbib}
\bibliography{strings,refs}

\begin{thebibliography}{10}

\bibitem{retkowski2025speech}
Fabian Retkowski, Maike Z{\"u}fle, Andreas Sudmann, Dinah Pfau, Jan Niehues, and Alexander Waibel,
\newblock ``From speech to summary: A comprehensive survey of speech summarization,''
\newblock {\em arXiv preprint arXiv:2504.08024}, 2025.

\bibitem{matsuura2023leveraging}
Kohei Matsuura, Takanori Ashihara, Takafumi Moriya, Tomohiro Tanaka, et~al.,
\newblock ``Leveraging large text corpora for end-to-end speech summarization,''
\newblock in {\em ICASSP}. IEEE, 2023, pp. 1--5.

\bibitem{matsuura2023transfer}
Kohei Matsuura, Takanori Ashihara, Takafumi Moriya, Tomohiro Tanaka, et~al.,
\newblock ``Transfer learning from pre-trained language models improves end-to-end speech summarization,''
\newblock {\em arXiv preprint arXiv:2306.04233}, 2023.

\bibitem{sharma2022end}
Roshan Sharma, Shruti Palaskar, Alan~W Black, and Florian Metze,
\newblock ``End-to-end speech summarization using restricted self-attention,''
\newblock in {\em ICASSP}. IEEE, 2022, pp. 8072--8076.

\bibitem{shang2024end}
Hengchao Shang, Zongyao Li, Jiaxin Guo, Shaojun Li, et~al.,
\newblock ``An end-to-end speech summarization using large language model,''
\newblock {\em arXiv preprint arXiv:2407.02005}, 2024.

\bibitem{kang2024prompting}
Wonjune Kang and Deb Roy,
\newblock ``Prompting large language models with audio for general-purpose speech summarization,''
\newblock {\em arXiv preprint arXiv:2406.05968}, 2024.

\bibitem{eom2024squba}
SooHwan Eom, Jay Shim, Eunseop Yoon, Hee~Suk Yoon, Hyeonmok Ko, Mark~A Hasegawa-Johnson, and Chang~D Yoo,
\newblock ``Squba: Speech mamba language model with querying-attention for efficient summarization,''
\newblock .

\bibitem{hurst2024gpt}
Aaron Hurst, Adam Lerer, Adam~P Goucher, Adam Perelman, et~al.,
\newblock ``Gpt-4o system card,''
\newblock {\em arXiv preprint arXiv:2410.21276}, 2024.

\bibitem{comanici2025gemini}
Gheorghe Comanici, Eric Bieber, Mike Schaekermann, Ice Pasupat, et~al.,
\newblock ``Gemini 2.5: Pushing the frontier with advanced reasoning, multimodality, long context, and next generation agentic capabilities,''
\newblock {\em arXiv preprint arXiv:2507.06261}, 2025.

\bibitem{chu2024qwen2}
Yunfei Chu, Jin Xu, Qian Yang, Haojie Wei, Xipin Wei, Zhifang Guo, Yichong Leng, Yuanjun Lv, Jinzheng He, Junyang Lin, et~al.,
\newblock ``Qwen2-audio technical report,''
\newblock {\em arXiv preprint arXiv:2407.10759}, 2024.

\bibitem{abouelenin2025phi}
Abdelrahman Abouelenin, Atabak Ashfaq, Adam Atkinson, Hany Awadalla, et~al.,
\newblock ``Phi-4-mini technical report: Compact yet powerful multimodal language models via mixture-of-loras,''
\newblock {\em arXiv preprint arXiv:2503.01743}, 2025.

\bibitem{gu2023minillm}
Yuxian Gu, Li~Dong, Furu Wei, and Minlie Huang,
\newblock ``Minillm: Knowledge distillation of large language models,''
\newblock {\em arXiv preprint arXiv:2306.08543}, 2023.

\bibitem{agarwal2024policy}
Rishabh Agarwal, Nino Vieillard, Yongchao Zhou, Piotr Stanczyk, Sabela~Ramos Garea, Matthieu Geist, and Olivier Bachem,
\newblock ``On-policy distillation of language models: Learning from self-generated mistakes,''
\newblock in {\em ICLR}, 2024.

\bibitem{cao2025llm}
Yihan Cao and Yanbin Kang,
\newblock ``On llm knowledge distillation-a comparison between forward kl and reverse kl,''
\newblock in {\em The Fourth Blogpost Track at ICLR 2025}.

\bibitem{rafailov2023direct}
Rafael Rafailov, Archit Sharma, Eric Mitchell, Christopher~D Manning, Stefano Ermon, and Chelsea Finn,
\newblock ``Direct preference optimization: Your language model is secretly a reward model,''
\newblock {\em NeuIPS}, vol. 36, pp. 53728--53741, 2023.

\bibitem{li2014learning}
Jinyu Li, Rui Zhao, Jui-Ting Huang, and Yifan Gong,
\newblock ``Learning small-size {DNN} with output-distribution-based criteria,''
\newblock in {\em Proc. Interspeech}, 2014, pp. 1910--1914.

\bibitem{hinton2015distilling}
Geoffrey Hinton, Oriol Vinyals, and Jeff Dean,
\newblock ``Distilling the knowledge in a neural network,''
\newblock {\em arXiv preprint arXiv:1503.02531}, 2015.

\bibitem{sutton1999policy}
Richard~S Sutton, David McAllester, Satinder Singh, and Yishay Mansour,
\newblock ``Policy gradient methods for reinforcement learning with function approximation,''
\newblock {\em Advances in neural information processing systems}, vol. 12, 1999.

\bibitem{weng2024rewardhack}
Lilian Weng,
\newblock ``Reward hacking in reinforcement learning.,''
\newblock {\em lilianweng.github.io}, Nov 2024.

\bibitem{sheng2025hybridflow}
Guangming Sheng, Chi Zhang, Zilingfeng Ye, Xibin Wu, Wang Zhang, Ru~Zhang, Yanghua Peng, Haibin Lin, and Chuan Wu,
\newblock ``Hybridflow: A flexible and efficient rlhf framework,''
\newblock in {\em EuroSys}, 2025, pp. 1279--1297.

\bibitem{kwon2023efficient}
Woosuk Kwon, Zhuohan Li, Siyuan Zhuang, Sheng, et~al.,
\newblock ``Efficient memory management for large language model serving with pagedattention,''
\newblock in {\em Proceedings of the 29th symposium on operating systems principles}, 2023, pp. 611--626.

\bibitem{carletta2005ami}
Jean Carletta, Simone Ashby, Sebastien Bourban, Mike Flynn, et~al.,
\newblock ``The {AMI} meeting corpus: A pre-announcement,''
\newblock in {\em International workshop on machine learning for multimodal interaction}. Springer, 2005, pp. 28--39.

\bibitem{chen2024floras}
William Chen, Brian Yan, Chih-Chen Chen, and Shinji Watanabe,
\newblock ``Floras 50: A massively multilingual multitask benchmark for long-form conversational speech,''
\newblock in {\em 2024 SLT}. IEEE, 2024, pp. 891--898.

\end{thebibliography}

\end{document}